\documentstyle[12pt,fleqn]{article}
\author{L.~Didukh and O.~Kramar  \\
{\small \it Ternopil State Technical University, Department of Physics}
\protect\\ {\small \it 56 Rus'ka Str., Ternopil UA--46001, Ukraine}\\
{\small \it E-mail: didukh@tu.edu.te.ua}}
\date{}

\title{On metallic ferromagnetism of a generalized Hubbard model with
correlated hopping}		
\begin{document}
\maketitle

\begin{abstract}
In the paper a possibility of metallic ferromagnetic state realization
in a generalized Hubbard model with more complete accounting of
electron-electron interactions, in particular, the correlated hopping 
and exchange interaction integrals is investigated.
Recently obtained by means of mean-field approximation
single electron energy spectrum is used for the description  
of finite temperature properties of the system.
In the paper the expression for the critical temperature of
ferromagnet-paramagnet transition is found, the behaviour
of temperature dependencies of magnetization and paramagnetic 
susceptibility is analyzed.
Taking into account the correlated hopping allows to explain some peculiarity
of ferromagnetic behaviour of transition metals, their alloys and
compounds.

PACS number(s): 71.10.Fd, 71.30.+h, 71.27.+a
\end{abstract}

\section{Introduction}  
The important puzzle in the explanation of ferromagnetism in the system 
of single band electrons is proper taking into account the Coulomb 
correlation between electrons. 
The Hubbard model~\cite{hub,kan,gutz}, which describes the single 
non-degenerate electron band with local Coulomb interaction, is 
incapable to describe ferromagnetism and requires 
the generalization. It is naturally to generalize the model Hamiltonian
by accounting  other matrix elements of electron-electron interaction 
(in addition to the intra-atomic Coulomb repulsion) and to consider 
the ferromagnetism in the generalized Hubbard model (for review on this
problem see Refs.~\cite{tas,dieter,irkh}).

Note that last time the problem of metallic 
ferromagnetism in the single band Hubbard model and their generalizations
has attracted much attention of researchers in a series of papers by means 
of the Gutzwiller variational wave function 
approximation~\cite{kol&voll,nie&zhou}, 
the dynamical mean-field theory~\cite{voll,wahl},
the spectral density approximation~\cite{nolt},
the exact diagonalization method~\cite{amad} and mean-field 
theory~\cite{hir_t0,hir_t,hir_ex}.
In the papers of Hirsch~\cite{hir_t0,hir_t,hir_ex} 
the generalization of Stoner-Wolfarth theory~\cite{ston,wolf} was carried out. 
The Stoner-Wolfarth theory has been used for the description
of band electron magnetism (denote in this connection, that the calculations
in this theory essentially depend on the shape and peculiarities of the 
density of states, in particular, it is known that the incomplete 
ferromagnetism in this model is absent if the density of state 
is rectangular~\cite{wolf}).
The considered generalization~\cite{hir_t0,hir_t,hir_ex} has shown
that the inter-atomic exchange interaction
plays the key role for the obtaining of ferromagnetic state with partially
polarization. Using the mean-field theory Hirsch has obtained the
condition of ferromagnetic state realization, nonmonotonic behaviour of
the concentration dependence of magnetization, the expression for the
Curie temperature, temperature dependencies of magnetization and 
magnetic susceptibility. 
The authors of Ref.~\cite{nie&zhou} using local approximation developed
from the Gutzviller wave function method also suggest 
(in agreement with~\cite{hir_t0}) the importance of inter-atomic exchange 
interaction for the stabilization of ferromagnetic state with partial spin
polarization.
In the Ref.~\cite{ivan}  the opposite case of strong
correlation has been studied and the criterion of ferromagnetism and
magnetization of the system in ground state have been derived.

At our point of view it is in principle necessary to consider in addition 
to the inter-atomic exchange interaction the matrix elements that describe
the correlated hopping of electrons~\cite{did} for the explanation
of ferromagnetism in single band model. 
Taking into account the correlated hoppind has alloved to explain some 
peculiarities of metal-insulator transition in single band  
model~\cite{cmp,dh_ltp} and double orbitally degenerate model~\cite{prb}.
In this model an electron hopping from one site to another is correlated 
both by the occupation of the sites involved in the hopping process and the 
occupation of the nearest-neighbour sites.
Taking into account the correlated hopping allows to obtain the additional
mechanism~\cite{dks_cmp,cm00} for the stabilization of ferromagnetic 
ordering.

The importance of correlated hopping for understanding of the metallic
ferromagnetism  was discussed in Ref.~\cite{amad}, where the results,
obtained by means of exact diagonalization and mean-field theory
for small one-dimensional chain of atoms were compared.  
It has been shown that in strong coupling regime
correlated hopping favours ferromagnetism stronger for electron
concentration $n>1$ than for $n<1$ (the reverse situation occurs at weak
interactions~\cite{amad}); this result agrees with conclusions
of Ref.~\cite{did}.
Using the Gutzviller approach the authors of Ref.~\cite{kol&voll} have
shown, that the correlated hopping favour the ferromagnetic ordering 
around the point of half-filling.

The magnetic properties of the system at nonzero temperature in the case of
strong and intermediate electron correlations was analyzed  
in a series of papers~\cite{voll,wahl,nolt,pott}. 
However the important question concerning the influence of correlated hopping
on the Curie temperature, on the behaviour of magnetization and magnetic 
susceptibility at nonzero temperature is not yet considered.

The main goal of the article is to show the principle necessity
to take into account both correlated hopping of electrons and
inter-atomic exchange for the correct description of metallic
ferromagnetism. 
Note that the application of mean-field theory leads to the 
overestimation of ferromagnetic tendencies and gives the results,
which agree with experiment only qualitatively.
At the same time it is interesting to investigate the case 
of weak electron correlations and to define the mechanismes which
are responsible for the ferromagnetic ordering.
The mean-field analyzis in spite of its limitation
allows to reproduce the behaviour of the magnetic moment and
Curie temperature at the change of electron concentration, the  correct
temperature dependence of magnetic moment of the system.
Besides we believe that taking into account the correlated hopping
allows to avoid the overestimation of Curie temperature. 

The paper has the following structure. In the Sec.~2 the Hamiltonian of
the generalized Hubbard model with correlated hopping and inter-atomic 
exchange interaction is written and the single electron energy spectrum
obtained in the case weak electron interactions by means of Green function
technique is analyzed.   
In the Sec.~3 the ground state properties of the system is considered,
the equation on the system critical parameters and expression 
for the magnetization are derived. 
In the Sec.~4 the finite temperature properties of the system are considered,
the expression for the Curie temperature, temperature dependencies of
magnetization and magnetic susceptibility are obtained.
The Sec.~5 is devoted to the conclusions.

\section{The model Hamiltonian and single electron energy spectrum}
Let us write the Hamiltonian, proposed in Refs.~\cite{cmp,dks_cmp}, 
generalized by taking into account weak magnetic field
\begin{eqnarray} \label{Ham}
H=&-&\mu \sum_{i\sigma}a_{i\sigma}^{+}a_{i\sigma}
+{\sum_{ij\sigma}}' t_{ij}(n)a_{i\sigma}^{+}a_{j\sigma}
+{\sum_{ij\sigma}}'\left(T_2(ij)a_{i\sigma}^{+}a_{j\sigma}n_{i\bar{\sigma}}
+h.c.\right)
\nonumber\\ 
&+&U\sum_{i}n_{i\uparrow}n_{i\downarrow}
+{J\over 2}{\sum_{ij\sigma \sigma^{'}}}' a_{i\sigma}^{+}
a_{j\sigma{'}}^{+}a_{i\sigma{'}}a_{j\sigma}-h
\sum_{i}\left(n_{i\uparrow}-n_{i\downarrow}\right),
\end{eqnarray}
where $a_{i\sigma}^{+}$, $a_{i\sigma}$ are creation
and destruction operators of electron on site $i$, 
$\sigma =\uparrow, \downarrow$, $n_{i\sigma}=a_{i\sigma}^{+}a_{i\sigma}$,
 $n_{i\sigma}=a_{i\sigma}^{+}a_{i\sigma}$ is the operator of 
number of electrons with spin $\sigma$ on site $i$,
$n=\langle n_{i\uparrow}+n_{i\downarrow}\rangle$,
$\mu$ is the chemical potential, 
$t_{ij}(n)=
t_{ij}+n
T_{1}(ij)$
is the effective hopping integral of an electron 
from site $j$ to site $i$,
$t_{ij}$ is the band hopping integral an electron 
from site $j$ to site $i$,
$T_{1}(ij)$ and $T_2(ij)$ is the parameters of correlated hopping of electrons,
$U$ is the intra-atomic Coulomb repulsion, 
$J$ is the exchange integral for the nearest neighbours
and $h$ is external magnetic field (the units of $h$ use such
the magnetic moment per electron is unity).
The prime on the sums in Eq.~(\ref{Ham}) signifies that $i\neq{j}$. 
The concentration dependence of the effective hopping integral
$t_{ij}(n)$ is caused by the correlated hopping~\cite{cmp}
of electrons.

The peculiarities of the model described by the Hamiltonian~(\ref{Ham}) 
are taking into account the influence of site occupation on hopping
process (correlated hopping) and the direct exchange
interaction between electrons in neighbour sites.
In this model an electron hopping from one site to another is correlated 
both by the occupation of the sites involved in the hopping process 
and the occupation of the nearest-neighbour sites (in contrast to similar 
generalized Hubbard models). 

To characterize the value of the correlated hopping we introduce dimensionless
parameters $\tau_1=\frac{T_1(ij)}{|t_{ij}|}$, 
$\tau_2=\frac{T_2(ij)}{|t_{ij}|}$ which are independent on the number of
site.

In terms of Green functions method using the mean-field approximation
we have recently
obtained~\cite{cm00} the single electron energy spectrum, 
which in the external magnetic field has the following form
\begin{eqnarray}\label{spectr}
&& E_\sigma({\bf k})=-\mu+\beta_\sigma+n_{\bar{\sigma}}U-zn_{\sigma}J
-h \eta_\sigma+
t(n\sigma)
\gamma({\bf k});
\end{eqnarray}
here the spin-dependent shift of the band centers is
\begin{eqnarray}
&& \beta_\sigma={2 \over N}\sum_{ij} T_2(ij) \langle 
a_{i\bar{\sigma}}^{+}a_{j\bar{\sigma}}\rangle,
\end{eqnarray}
z is the number of nearest neighbours to a site,
for the spin~$\sigma=\uparrow(\downarrow)$ we have  
$\eta_\sigma=1~(-1)$,
$\gamma({\bf k})=
\sum\limits_{{\bf R}}e^{i{\bf kR}}$
(the sum goes over the nearest neighbours to a site),
and the spin and concentration dependent hopping integral is
\begin{eqnarray}\label{tef}
&& t(n\sigma)=(1-\tau_1n-2 \tau_2 n_{\bar{\sigma}}-{zJ \over w} \sum_{\sigma'}
A_{\sigma'})t=\alpha_\sigma t,
\end{eqnarray}
where $w=z|t|$ is the half of band width, $t$ is the hopping integrals
between nearest neighbour sites and  
\begin{eqnarray}
A_{\sigma'}={1 \over N}
\sum_{ij} \left(-\frac{t_{ij}}{w} \langle 
a_{i\sigma'}^{+}a_{j\sigma'}\rangle\right).
\end{eqnarray}
The energy spectrum~(\ref{spectr}) will be used in the next sections
for the description of the model properties in the ground state and
at the finite temperature.

\section{The ground state properties of the model}
In general the occupation number and the magnetization 
are expressed respectively as
\begin{eqnarray}
\label{n}
n=n_\uparrow+n_\downarrow=\int\limits_{-w}^{w}\rho(\epsilon)
\left[ f(E_\uparrow(\epsilon))+f(E_\downarrow(\epsilon))\right] d\epsilon;
\end{eqnarray}
\begin{eqnarray}
\label{m}
m=n_\uparrow-n_\downarrow=\int\limits_{-w}^{w}\rho(\epsilon)
\left[ f(E_\uparrow(\epsilon))-f(E_\downarrow(\epsilon))\right] d\epsilon,
\end{eqnarray}
where $\rho(\epsilon)$ is the density of states, $f(E_\sigma(\epsilon))$ 
is the Fermi function. 

Generalizing the recently obtained result~\cite{cm00} 
(for the rectangular density of states) in the case of
weak external magnetic field one can obtain for
the correlation function $A_\sigma$ and
the shift of the center of $\sigma$-subband $\beta_{\sigma}$ 
\begin{eqnarray}
\label{A}
A_{\sigma}=n_\sigma(1-n_\sigma),
\end{eqnarray}
\begin{eqnarray}
\label{beta}
\beta_\sigma=2w\tau_2A_{\sigma}=2w\tau_2n_\sigma(1-n_\sigma),
\end{eqnarray}
where the concentration of electrons with spin $\sigma$  
\begin{eqnarray}
n_{\sigma}=\frac{\varepsilon_{\sigma}+w}{2w},
\end{eqnarray}
here $\varepsilon_{\sigma}=\frac{\mu_{\sigma}}{\alpha_{\sigma}}$
is the solution of the equation  $E_{\sigma}(\epsilon)=0$,
where $\mu_{\sigma}=\mu-\beta_{\sigma}+zn_{\sigma}J-n_{\bar\sigma}U
+h \eta_\sigma$
and $\alpha_{\sigma}=1-\tau_1n-2\tau_2 n_{\bar\sigma}
-{zJ\over w}\sum\limits_{\sigma'}A_{\sigma'}$.

On basis of expressions~(\ref{n})-(\ref{m}) using~(\ref{A})-(\ref{beta})  
one can obtain the equation on the system critical parameters 
\begin{eqnarray}
\label{umova_m}
{zJ \over 2w}\left(1+n(2-n)-m^2\right) =
1-n\tau_1-\tau_2(2-n)-{U \over 2w}-{h \over w}.
\end{eqnarray}
The condition for onset of ferromagnetic ordering is obtained
(replacing the sign of equality by sign of inequality), when $m=0$, and
the condition for the ferromagnetic state with full spin
polarization if one put $m=n$ in the expression (\ref{umova_m}). 
The peculiar distinction of the obtained expression from the similar
condition, obtained in Refs.~\cite{hir_t0,hir_t,hir_ex,nie&zhou}
for the various generalizations of the Hubbard model is the presence
of correlated hopping, which can essentially modify the properties
of the model.
Taking into account correlated hopping leads to the appearance
of a peculiar kinetic mechanism of ferromagnetic ordering stabilization. 
This mechanism is caused by the presence of the spin-dependent shift of
the spin subband centers being the consequence of correlated hopping
(which are analogous to the shift of subband centers in consequence of 
inter-atomic direct exchange interaction).

The spontaneous magnetization (in the absence of magnetic field) of the
system is found from the equation (\ref{umova_m}) as
\begin{eqnarray}
\label{m0} 
m^{GS}=\sqrt{1+n(2-n)-{1-U/2w-\tau_1 n-\tau_2 (2-n)
\over zJ/2w}},
\end{eqnarray}
which is valid in the case $J>0$ (when $J=0$ only the transition from
paramagnetic to fully polarized ferromagnetic state with $m^{GS}=n$
occurs).
If the calculated magnetization $m^{GS}>n$ it is necesarry to put $m^{GS}=n$.

The influence of correlated hopping on the properties of the system
is illustrated on the Figs.~1,~2. 
On the Fig.~1 the dependence of the critical value of exchange integral
at which the ferromagnetic ordering occurs is plotted (at the absence of
magnetic field) as a function of band filling at various values of correlated
hopping parameters $\tau_1$, $\tau_2$ and intra-atomic Coulomb interaction 
$U/w$. Solid curves correspond to the onset of spin polarization,
dashed curves correspond to the fully polarized ferromagnetic state
(the area below the solid line is paramagnetic, above the dashed line - fully
polarized ferromagnetic, between lines - partially polarized ferromagnetic).  
The similar phase diagrams have been obtained in the Refs.~\cite{hir_t0,
hir_ex}, but in those works the correlated hopping was not considered.
Note that the presence of correlated hopping (curves~2,~3 on the Fig.~1)
leads to the essential change of phase boundary of paramagnet-ferromagnet
transition, in particular, to the shift of the minimum point, namely to the
inequivalence of the cases $n<1$ and $n>1$.

On the Fig.~2 the dependence of ground state magnetization $m^{GS}$
on the electron concentration $n$ at the fixed values of exchange and 
Coulomb interactions, as  well as at different values of correlated hopping
parameters is plotted.
In Refs.~\cite{hir_t0,hir_ex} the concentration dependencies
of magnetization in the framework of Hubbard model with inter-atomic exchange
were obtained, but they are symmetrical respecting the point of half-filling.
In the Ref.~\cite{ivan} (for the case $U \rightarrow \infty$) the similar
$m^{GS}(n)$ dependence was obtained.

As mentioned above, the asymmetry of cases $n<1$ and $n>1$ is obsereved;
in particular, the increase of parameter $\tau_1$ leads to the shift
of the ferromagnetic area in a region of larger electron concentration $n$,
the increase of parameter $\tau_2$ - in a region of smaller $n$. 
Note also that taking into account the correlated hopping considerably
enriches the set of curves $m^{GS}(n)$.
The obtained concentration dependencies of magnetization allow qualitatively
describe the experimental curves for the binary ferromagnetic alloys
of transition metals Fe, Co, Ni (Slater-Pouling's curves~\cite{gau}).

\section{The properties of the model at nonzero temperature}
At nonzero temperature and rectangular density of states the concentration
of electrons with fixed spin direction is expressed as
\begin{eqnarray}
n_\sigma=1-{\Theta \over {2w\alpha_\sigma}}
 ln \left( \frac{1+exp(E_\sigma(w))}{1+exp(E_\sigma(-w))}\right).
\end{eqnarray}
Using this expression one can obtain the equation for the magnetization
\begin{eqnarray}
exp\left(-{mJ_{eff}w \over \Theta}\right)={{sh\left( {(1-n_\uparrow)
\alpha_\uparrow w \over \Theta }\right) sh\left( {n_\downarrow
\alpha_\downarrow w \over \Theta }\right)}\over{sh\left( {(1-n_\downarrow)
\alpha_\downarrow w \over \Theta }\right) sh\left( {n_\uparrow
\alpha_\uparrow w \over \Theta }\right)}},
\end{eqnarray}
where $J_{eff}=zJ/w+U/w+2\tau_2(1-n)$.
To obtain the temperature dependence of magnetization it is necessary
to apply the numeric methods forasmuch the last equation can not be
solved analitically.
The numeric calculations show that the results could be approximately 
expressed using the approach proposed in Ref.~\cite{hir_t}.  
The critical temperature of transition from paramagnetic to 
ferromagnetic metallic state (the Curie temperature) can be obtained
by expanding~(\ref{m}) at lowest order of $m \to 0$. 
\begin{eqnarray}
1&=&{1 \over 2w}\frac{U+zJ-2w\tau_2(n-1)}{\alpha^*} \int \limits_{-w\alpha^*}
^{w\alpha^*}\left[ -{\partial f(x-\mu^*) \over \partial x} \right] dx
\nonumber\\
&-&
\frac{\tau_2}{w(\alpha^*)^2}\int \limits_{-w\alpha^*}
^{w\alpha^*} \left[-{\partial f(x-\mu^*) \over \partial x}\right]xdx,
\end{eqnarray}
here $x=\alpha^*\epsilon$, $\alpha^*=\alpha_\sigma|_{m=0}$ 
and $\mu^*=\mu_\sigma|_{m=0}$.

At low temperature one can approximately write
\begin{eqnarray}
\label{umova_t}
1=\frac{U/2w+zJ/2w-2w\tau_2(n-1)}
{1-(\tau_1+\tau_2)n -2A^*(\Theta_C)zJ/w},
\end{eqnarray}
where $A^*(\Theta_C)=A_\sigma(\Theta)|_{m=0}$, $\Theta=k_BT$, $k_B$ is 
Boltzmann constant. In the case of zero temperature the last expression
reproduces the criterion of the paramagnet-ferromagnet transition,
obtained in the recent paper~\cite{cm00}. 

To find $A_\sigma(\Theta)$ one can apply the Sommerfeld 
expansion~\cite{ashkr} (at low temperature) 
\begin{eqnarray}
\label{A_t}
A_\sigma(\Theta)&=&\int \limits_{-w}^w \left[-{\epsilon \over 2w^2}\right]
{d\epsilon  \over exp(E_\sigma(\epsilon)/\Theta)+1}
\\
&\simeq&n_\sigma(1-n_\sigma)-{\pi^2 \over 3}
 \left({\Theta \over 2w}\right)^2
\frac{1}{\left[ 1-\tau_1n-2\tau n_{\bar \sigma}-(A_\uparrow(\Theta)+
A_\downarrow(\Theta))zJ/w \right]^2}.
\nonumber
\end{eqnarray}
In the Curie point (when the magnetic moment $m \rightarrow 0$ and $n_\sigma=n/2$)
the last equation is written as
\begin{eqnarray}
\label{A(T_c)}
A^*(\Theta_C)={n(2-n) \over 4}-{2\pi^2 \over 3}
 \left({\Theta_C \over 2w}\right)^2
\frac{1}{\left[ 1-(\tau_1+\tau_2)n-2A^*(\Theta_C)zJ/w \right]^2}.
\end{eqnarray}
Solving the system of equation~(\ref{umova_t}) and~(\ref{A(T_c)}) we can
express the Curie temperature as function of model parameters 
\begin{eqnarray}
{\Theta_C \over 2w}&=&\sqrt{1+n(2-n)-{1-U/2w-\tau_1 n-\tau_2 (2-n)
\over zJ/2w}}\times
\nonumber\\
&&
\sqrt{3 \over 2\pi^2}
\left({U \over 2w}+{zJ \over 2w}-2\tau_2(n-1) \right).
\end{eqnarray}
Taking into account the expression~(\ref{m0}) we finally obtain for
the Curie temperature
\begin{eqnarray}
\label{T_c}
{\Theta_C \over 2w}&=&\sqrt{3 \over 2\pi^2}
\left({U \over 2w}+{zJ \over 2w}-2\tau_2(n-1) \right)m^{GS}.
\end{eqnarray}
At zero correlated hopping the formula~(\ref{T_c}) reproduces the result
of Ref.~\cite{hir_t}. 
Note that the Curie temperature is tightly related (in the 
used approximation) to ground state magnetization of ferromagnet. 
The peculiarity of formula~(\ref{T_c}) is the presence of term related
to correlated hopping $\tau_2$ ($m^{GS}$ is also dependent on this parameter),
which in the case $n>1$ is negative and therefore lowers the Curie temperature.
This circumstance (at correspondent choice of correlated hopping parameters),
at our opinion allow to overcome the considerable defect of mean-field
treatment, namely the overestimation of the Curie temperature
(in this connection also see Ref.~\cite{hir_t}, where the correlated
hopping is not took into account). 
Let us put (as in the Ref.~\cite{hir_t}) the bandwidth of the $e_g$ states
in 3d ferromagnetic transition metals is approximately 2 eV. In our work
the band filling $n=1.2$ corresponds to Fe and for the values
of correlated hopping $\tau_1=0.15$, $\tau_2=0.2$ our theory would predict
the Curie temperature between 1000 and 1600 K depending on the values of 
$U/w$ and $zJ/w$ (these parameters are changed in the region from 0 to 0.4 and
from 0.37 to 0.14 respectively). The value of Curie temperature  at some
fixed values of intra-atomic Coulomb interaction and inter-atomic exchange
interaction agrees with experimental data even quantitatively.  

On the Fig.~3 the concentration dependence of the Curie temperature 
is plotted at various values of correlated hopping. 
The peculiarity of this dependence is the lowering of the Curie temperature
at the increase of carriers concentration.
Besides taking into account the correlated hopping causes the asymmetry
of curve respecting the point of half-filling, that allows to explain 
qualitatively the larger Curie temperature in Co as compared to Fe.
Note for the comparison that this fact can not be explained in the 
Ref.~\cite{hir_t} without further comments (in particular, taking into
account the density of states with peculiarities). 
On basis of obtained expression for the Curie temperature one can
explain the peculiarities of Curie temperature behaviour in the binary alloys
of transition metals~\cite{gau}. It is reasonable in the framework of our
theory to interpret the paramagnet-ferromagnet transition in metallic phase
at the increase of temperature for the nonstoichiometric chalcogenide chrome 
spinel Cr$_{0.5}$Fe$_{0.5}$S (where the Curie temperature is of the  
order of 1000 K)~\cite{loseva}.

In principle, our theory allows to obtain the concentration
dependences of magnetization and Curie temperature, which are similar to
the experimentally observed in compounds  Fe$_{1-x}$Co$_x$S$_2$ 
and Co$_{1-x}$Ni$_x$S$_2$ with the change of 
electron concentration in $3d$-band~\cite{jarr}. In these crystals
the same subsystem of electrons is responsible both for conductivity
and for the localized magnetic moments formation.
Although these compounds are adequately described in the framework of double
orbitally degenerate model, even in terms of single band model
at some values of model parameters it is possible to obtain the concentration
dependencies of the Curie temperture and magnetic moment of system 
(see. Fig.~4) that agree qualitatively with experimentally 
observed~\cite{jarr}.
The plotted curves show that at choosen model parameters the values
of mentioned above quantities are reproduced accurately. 

The influence of model parameters on the critical temperature
is illustrated in the next figures. On the  Figs.~5 and 6 the dependencies 
of the Curie temperature on the parameter of intra-atomic Coulomb interaction
are plotted at half-filling (Fig.~5) and various values of band filling
(Fig.~6), the values of exchange integral and correlated hopping are fixed.
The plotted curves have a peculiarities: one can distinquish the area
of sharp increase of the Curie temperature at the increase of parameter $U/w$ 
(this values of model parameters correspond to the partial spin polarization 
of system) and the area, where the Curie temperature changes in proportion to 
$U/w$ (this values correspond to full spin polarization of system). Note that
the increase of inter-atomic exchange interaction leads to the extension
of partially polarized ferromagnetic area, and also to the decrease of critical
value of $U$, required for the development of saturated ferromagnetism.
The peculiarity of the dependence of the Curie temperature on the 
$U/w$ is the increase of critical value  $U/w$ at the increase 
of $n$. Note also that the changing of this critical value
(at choosen values of correlated hopping) in the case $n>1$
and at the increase of electron concentration is more pronounced than in the
case $n<1$. It is also interesting that depending on the magnitude of $U/w$
the system with electron concentration $n<1$ can have the larger value of
the Curie temperature than the system with $n>1$. However starting
from the critical point for the intra-atomic Coulomb interaction 
the situation becomes opposite: the Curie temperature is larger for the 
system with $n>1$. 
 
Next consider the behaviour of the magnetization of the system
at the change of temperature. To obtain the temperature dependence of
magnetization $m$ let us use the assumption (as in the Ref.~\cite{hir_t}), 
that the equation~(\ref{umova_t}) is also valid for the nonzero magnetization
and for the temperature lower than critical $\Theta_C$.
Than using~(\ref{A_t}) one can obtain the equaiton
\begin{eqnarray}
m={2\pi \over \sqrt{3}}\sqrt{\left({\Theta_C \over 2w}\right)^2
{1 \over (\alpha^*)^2}-{1 \over 2}\left({\Theta \over 2w}\right)^2
\left[{1 \over (\alpha^*+\tau_2m)^2}+{1 \over (\alpha^*-\tau_2m)^2}
\right]}.
\end{eqnarray}
The results of numeric calculations for $m$ is plotted on the Fig.~7. 
It should be noted that if the correlated hopping is not take into account, 
the last expression gives the analitical result of Ref.~\cite{hir_t}. 
In the band limit last expression reproduces the result of Ref.~\cite{blan}, 
where the similar treatment is applied to the completely itinerant carriers.

To find the magnetic susceptibility let us to take the deriviative from
magnetization (\ref{m}) with respect to the magnetic field 
\begin{eqnarray}
\chi(T)={\partial m(h) \over \partial h}\vert_{h\rightarrow 0,m 
\rightarrow 0}={1 \over 2w}{1 \over \alpha^*-
U/2w-zJ/2w+2\tau(n-1)}.
\end{eqnarray}
Using~(\ref{umova_t}) at temperatures close to $\Theta_C$
one can obtain 
\begin{eqnarray}
\chi(T)={6w[U/2w+zJ/2w-2\tau_2(n-1)]^2 \over 2 \pi^2 
\Theta_C(\Theta-\Theta_C)zJ/w}.
\end{eqnarray}
On the Fig.~7 the temperature dependencies of magnetization and inverse
magnetic susceptibility are plotted. The similar plot was obtained 
for the transition metals in the Ref.~\cite{kats} using the dynamical mean 
field theory with local density approximation. 
At temperatures higher than Curie temperature the  magnetic susceptibility
demonstrates Curie-Weiss-like behaviour, intrinsic for the ferromagnetic
3d-metals and their alloys.
The similar temperature dependence of magnetic moment in alloy
$MnSb_xAs_{1-x}$ at $x>0.07$ and the presence of magnetic field is 
observed~\cite{zavad}.
  
\section{Conclusions}
In this paper the ferromagnetic solution in a single band generalized Hubbard
model with correlated hopping is derived.
The peculiarity of the model is the consideration so-called off-diagonal
matrix elements of Coulomb interaction between electrons, in particular
the integrals of correlated hopping (which describe the influence of occupancy
of sites on the hopping process) and inter-atomic exchange interaction. 
The physical mechanisms that lead to the ferromagnetic state realization
are the band narrowing and shift of the spin subband centers, caused by
the exchange interaction and correlated hopping.

Taking into account the correlated hopping leads to the asymmetry of cases
$n<1$ and $n>1$ at the consideration the ferromagnetism in this model.
The increase of parameter $\tau_1$ leads to the shift
of the ferromagnetic area in a region of larger electron concentration $n$,
the increase of parameter $\tau_2$ -- in a region of smaller electron
concentration $n$.
The important consequence of the performed study is the conclusion that 
for the ferromagnetism realisation the case of more than half-filled 
band is more favourable than less than half-filled one;this maps the situation
transition metals and their alloys.

The concentration dependence of magnetization $m^{GS}$ qualitatively agrees
with experimental data for the transition 3d-metals and their alloys,
in particular, our results can explain the Slater-Pouling's curves~\cite{gau}
for the binary ferromagnetic alloys of transition metals Fe, Co, Ni with
other 3d-metals.

Taking into account the correlated hopping at the calculation of the Curie 
temperature allows to avoid the problem of the overestimation of critical
temperature which is typical for the standard mean-field treatmens.
The calculated Curie temperature is characterized by peculiarities of 
concentration dependence, in particular, the asymmetry (in consequance of
correlated hopping accounting) respecting to the point of half-filling.
This fact allows to explain the experimentally observed values of critical
temperature in the transition ferromagnetic metals.
Besides our results qualitatively reproduce the non-typical
concentration dependence Curie temperature in systems Fe$_{1-x}$Co$_x$S$_2$ and 
Co$_{1-x}$Ni$_x$S$_2$~\cite{jarr}.

In conclusion the correct taking into account of above mentioned 
matrix elements of electron-electron correlations allows to explain
the behaviour of ferromagnetic transition metals, their alloys and compounds
both in the case of ground state and nonzero temperatures.

\newpage

{\bf Figure captions}

Fig.1 The critical values of $zJ/w$
as a function of $n$ at fixed value of $U/w$.
Curves 1 correspond to $U/w=1$, $\tau_1=\tau_2=0$,
curves 2 - to $U/w=1.2$, $\tau_1=0$, $\tau_2=0.15$, 
curves 3 - to $U/w=1.3$, $\tau_1=0.15$, $\tau_2=0.1$.

Fig.2 The ground state magnetization $m^{GS}$ as a function of 
$n$ at ${U/w}=1.2$ and ${zJ/w}=0.1$.
Curve 1 corresponds to $\tau_1=\tau_2=0$,
curve 2 -- to $\tau_1=0$, $\tau_2=0.05$
curve 3 -- to $\tau_1=0.05$, $\tau_2=0$,
curve 4 -- to $\tau_1=\tau_2=0.05$.

Fig.3 The Curie temperature as a function of $n$ at values of 
$U/w=1$ and $\tau_1=0.05$, $\tau_2=0.1$.
Curves 1 corresponds to $zJ/w=0.4$, 
curves 2 corresponds to $zJ/w=0.5$,
curves 3 corresponds to $zJ/w=0.6$.

Fig.4 The dependencies of magnetization (solid curve) and
Curie temperature (dashed curve) on electron concentration
$n$ at values of 
$U/w=1.5$, $\tau_1=0$, $\tau_2=0.08$ and $zJ/w=0.2$

Fig.5 The Curie temperature as a function of $U/w$ at
half-filling. 
Curve 1 corresponds to $zJ/w=0.4$, $\tau_1=\tau_2=0$; 
curve 2 corresponds to $zJ/w=0.4$, $\tau_1=\tau_2=0.1$;
curve 3 corresponds to $zJ/w=0.8$, $\tau_1=\tau_2=0$;
curve 4 corresponds to $zJ/w=0.8$, $\tau_1=\tau_2=0.1$.

Fig.6 The Curie temperature as a function of $U/w$ at
$zJ/w=0.4$, $\tau_1=0.05$, $\tau_2=0.2$ and various
band fillings. 
Curves 1,~2,~3,~4 and~5 correspond to $n=0.5, 0.75, 1, 1.25$ and
$1.5$ respectively.

Fig.7 The dependencies of magnetization and inverse magnetic 
susceptibility on reduced temperature at $U/w=1$, $zJ/w=0.5$,
$\tau_1=0.05$ and $\tau_2=0.1$; the upper, middle and lower curves 
correspond to the case $n=1$,$n=0.5$ and $n=0.2$ respectively.

\begin{thebibliography}{100}
\bibitem{hub} J.~Hubbard, Proc. Roy. Soc. {\bf A276}, 238 (1963).
\bibitem{kan} J.~Kanamori, Prog. Theor. Phys.{\bf 30}, 275 (1963).
\bibitem{gutz} M.~C.~Gutzwiller, Phys. Rev. Lett. {\bf 10}, 159 (1963).
\bibitem{tas} H.~Tasaki, Progr. Theor. Phys. {\bf99}, 489 (1998).
\bibitem{dieter} D.~Vollhardt et al., Adv. in Sol. St. Phys. {\bf38}, 383 
(1999).
\bibitem{irkh} V.~Yu.~Irkhin, Yu.~P.~Irkhin, {\it Electronic structure
and physical properties of d- and f-transition metals and their compounds},
Chap.~4, cond-mat/9812072. 
\bibitem{kol&voll} M.~Kollar and D.~Vollhardt, Phys. Rev.~B {\bf63}, 045107 
(2001).
\bibitem{nie&zhou} H.-Q.~Nie and W.-Y.~Zhou, Phys. Rev.~B {\bf55}, 59 (1997). 
\bibitem{voll} D.~Vollhardt, N.~Bl\" umer, K.~Held, M.~Kollar, J.~Schlipf, 
and M.~Ulmke, Z. Phys. B {\bf103}, 283 (1997). 
\bibitem{wahl} J.~Wahle, N.~Bl\" umer, J.~Schlipf,  K.~Held, 
D.~Vollhardt, Phys. Rev. B {\bf58}, 12749 (1998).
\bibitem{nolt} T.~Herrman and W.~Nolting, J. Magn. Magn. Mater. {\bf170},
253 (1997).
\bibitem{pott} M.~Potthoff, T.~Herrman and W.~Nolting, Phys. Stat. Sol (b)
{\bf210}, 199 (1997).
\bibitem{amad} J.~C.~Amadon  and J.~E.~Hirsch, Phys. Rev.~B {\bf54}, 6364
(1996).
\bibitem{hir_t0} J.~E.~Hirsch, Phys. Rev.~B {\bf40}, 2354 (1989).
\bibitem{hir_t} J.~E.~Hirsch, Phys. Rev.~B {\bf40}, 9061 (1989).
\bibitem{hir_ex} J.~E.~Hirsch, Phys. Rev.~B {\bf59}, 6256 (1999).
\bibitem{ston} E.~C.~Stoner, Proc. Roy. Soc. {\bf A165}, 372 (1938).
\bibitem{wolf} E.~P.~Wohlfart, Phil. Mag. {\bf 42}, 374 (1951).
\bibitem{ivan} V.~Ivanov, Ukr. Phys. Journ. {\bf 36}, 751 (1991). 
\bibitem{cmp} L.~Didukh, Cond. Matt. Phys. {\bf1}, 125 (1998). 
\bibitem{dh_ltp} L.~Didukh and V.~Hankevych, Fiz. Nizk. Temp. {\bf 25}, 481
(1999) [Low Temp. Phys. {\bf 25}, 354 (1999)].
\bibitem{prb} L.~Didukh, Yu.~Skorenky, Yu.~Dovhopyaty, and V.~Hankevych 
Phys. Rev.~B {\bf61}, 7893 (2000).
\bibitem{dks_cmp} L.~Didukh, O.~Kramar and Yu.~Skorenky, Cond. Matt. Phys.
 {\bf4}, 101 (2001).
\bibitem{cm00} L.~Didukh, O.~Kramar and Yu.~Skorenky, cond-mat/0012402.
\bibitem{did} L.~Didukh, Sov. Phys. Solid State {\bf19}, 711 (1977). 
\bibitem{ashkr} N.~W.~Ashcroft, N.~D.~Mermin, {\it  Solid State Physics},
(Holt,~Rinehart and~Winston, New York, 1975), Chap.~1. 
\bibitem{gau} F.~Gautier, in {\em Magnetism of Metals and Alloys\/}, edited
by M.~Cyrot (North-Holland, Amsterdam, 1982), Chap.~1.
\bibitem{loseva} G.~V.~Loseva, et al., {\it  Metal-insulator transition in
sulphides of 3d-metals}, (Nauka, Novosibirsk, 1983), Chap.~5 (in Russian).
\bibitem{jarr} H.~S.~Jarrett, et al., Phys. Rev. Lett. {\bf21}, 617 (1965).
\bibitem{blan} J.~A.~Blanco  and J.~Pisonero, Eur. J. Phys.~B {\bf20}, 289
(1998).
\bibitem{kats} A.~T.~Lichtenstein and M.~I.~Katsnelson, cond-mat/0102297.
\bibitem{zavad} E.~A.~Zavadsky, V.~I.~Valkov, {\it  Magnetic Phase Transition},
(Naukova dumka, Kyiv, 1980), Chap.~3 (in Russian). 
\end{thebibliography}
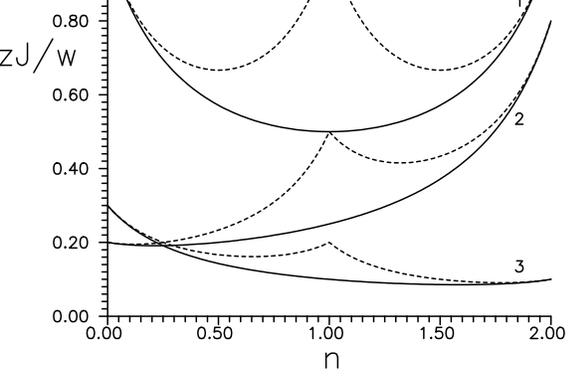
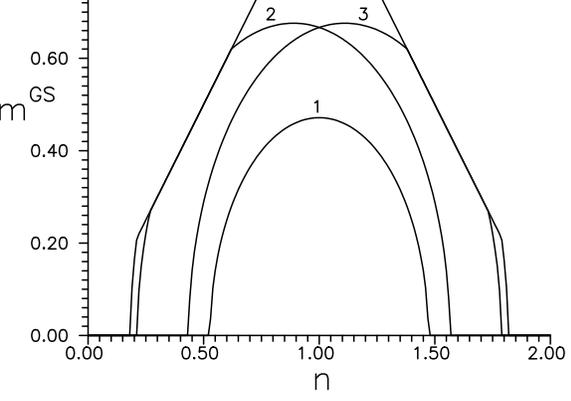
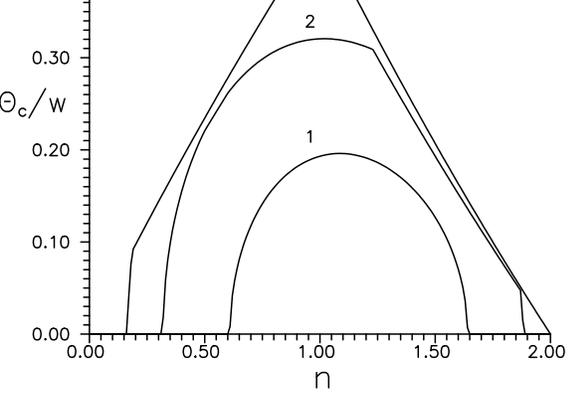
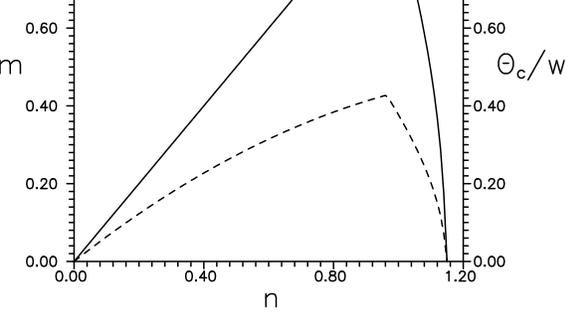
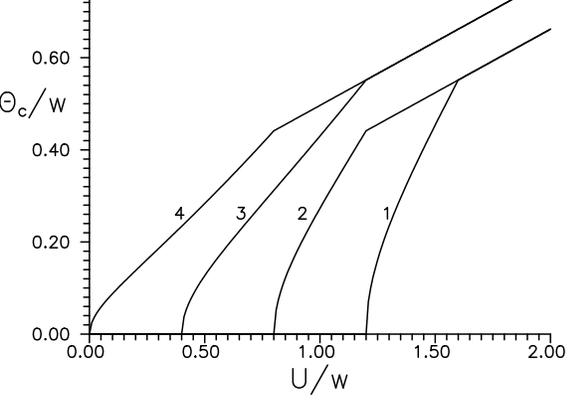
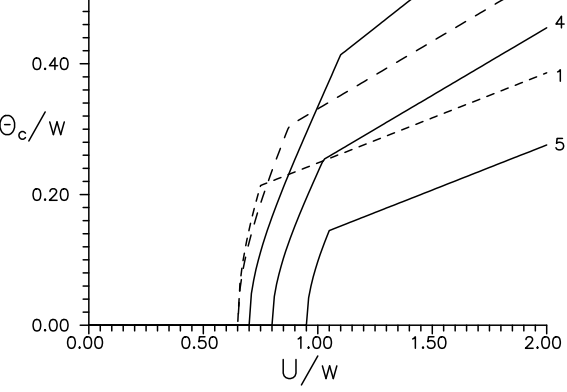
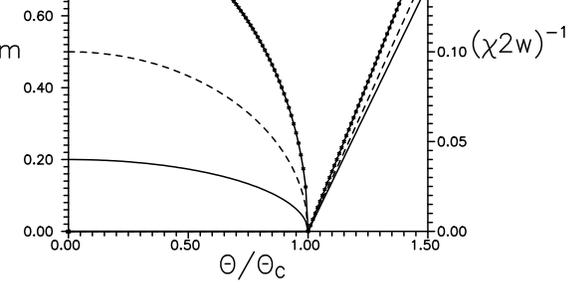
\end{document}